\documentclass{article}

\usepackage{hyperref}
\usepackage{cite}
\usepackage{amssymb,graphicx,amsmath,latexsym}
\newtheorem{theorem}{Theorem}
\newtheorem{definition}{Definition}
\newtheorem{corollary}{Corollary}

\def\eop{\hfill$\Box$}
\def\proof{{\em Proof: }}

\def\C{{\mathbb C}}

\begin{document}

\title{Graphs whose normalized Laplacian matrices are separable as density matrices in quantum mechanics}
\author{Chai Wah Wu\footnote{e-mail: chaiwahwu@member.ams.org}\\IBM T. J. Watson Research Center\\P. O. Box 218, Yorktown Heights, NY 10598, USA.}
\date{July 23, 2014}
\maketitle

\begin{abstract}
Recently normalized Laplacian matrices of graphs are studied as
density matrices in quantum mechanics.  Separability and entanglement of density matrices are important properties as they determine the nonclassical behavior in quantum systems.  In this note we look at the graphs whose normalized Laplacian matrices are separable or entangled.  In particular, we show that the number of such graphs is related to the number of $0$-$1$ matrices that are line sum symmetric and to the number of graphs with at least one vertex of degree $1$.
\end{abstract}

\section{Introduction}
Applications of quantum mechanics in information technology such as
quantum teleportation, quantum cryptography and quantum computing
\cite{nielsen-quantum-2002} lead to much recent interest in
studying entanglement in quantum systems.  One important problem is to
determine whether a given state operator is entangled or not.  This is
especially difficult for mixed state operators.  In
Refs. \cite{braunstein:laplacian:2006,braunstein:laplacian_graph:2006,wu:separable:2006,wang:tripartite:2007,wu:multipartite:2009,wu:separable:2010}, normalized Laplacian matrices
of graphs are considered as density matrices, and their entanglement
properties are studied.  The reason for studying this subclass of
density matrices is that simpler and stronger conditions for
entanglement and separability can be found and graph theory may shed light on the entanglement properties of state operators. In this note, we continue this study and determine the number of graphs that result in separable or entangled density matrices.

\section{Density matrices, separability, and partial transpose}
A state of a finite dimensional quantum
mechanical system is described by a state operator or a density matrix
$\rho$ acting on $\C^n$ which is Hermitian and positive semidefinite
with unit trace.  A state operator is called a {\em pure} state if 
it has rank one.  Otherwise the state operator is {\em mixed}.  An $n$ by
$n$ density matrix $\rho$ is separable in $\C^p\otimes \C^q$ with
$n=pq$ if it can be written as $\sum_{i} c_i \rho_i \otimes
\eta_i$ where $\rho_i$ are $p$ by $p$ density matrices and $\eta_i$
are $q$ by $q$ density matrices with $\sum_i c_i = 1$ and $c_i\geq
0$.\footnote{This definition can be extended to composite systems of
multiple states, but here we only consider decomposition into the
tensor product of two component states.}  A density matrix that is not
separable is called entangled.  Entangled states are necessary to
invoke behavior that can not be explained using classical physics and
enable novel applications.

We denote the $(i,j)$-th element of a matrix $A$ as $A_{ij}$.
Let $f$ be the canonical bijection between 
$\{1,\dots, p\}\times \{1,\dots,q\}$ and $\{1,\dots, pq\}$:
$f(i,j)  = (i-1)q+j$.  For a $pq$ by $pq$ matrix $A$,
if $f(i,j) = k$ and $f(i_2,j_2) = l$, we can write
$A_{kl}$ as $A_{(i,j)(i_2,j_2)}$.
\begin{definition}
The $(p,q)$-partial transpose $A^{PT}$ of an $n$ by $n$ matrix $A$,
where $n=pq$, is given by:
\[ A_{(i,j)(k,l)}^{PT} = A_{(i,l)(k,j)} \]
\end{definition}
We remove the prefix ``$(p,q)$'' if $p$ and $q$ are clear
from context.  In matrix form, the partial transpose is constructed by decomposing $A$ into $p^2$ blocks
\begin{equation}\label{eqn:A}
 A = \left(\begin{array}{cccc}
A^{1,1} & A^{1,2} &\cdots & A^{1,p} \\
A^{2,1} & A^{2,2} & \cdots & A^{2,p} \\
\vdots  & \vdots  &  & \vdots \\
A^{p,1} & A^{p,2} & \cdots      & A^{p,p} \end{array}\right)
\end{equation}
where each $A^{i,j}$ is a $q$ by $q$ matrix, and
$A^{PT}$ is given by:
\begin{equation}\label{eqn:Apt}
 A^{PT} = \left(\begin{array}{cccc}
(A^{1,1})^T & (A^{1,2})^T &\cdots & (A^{1,p})^T \\
(A^{2,1})^T & (A^{2,2})^T & \cdots & (A^{2,p})^T \\
\vdots  & \vdots & & \vdots \\
(A^{p,1})^T & (A^{p,2})^T & \cdots      & (A^{p,p})^T \end{array}\right)
\end{equation}

\subsection{Necessary conditions for separability of density matrices}
It is clear that if $A$ is Hermitian, then so is $A^{PT}$.
Peres \cite{peres:separability:1996} introduced the following necessary condition for
separability:
\begin{theorem} \label{thm:peres}
If a density matrix $\rho$ is separable, 
then $\rho^{PT}$ is positive semidefinite, i.e.
$\rho^{PT}$ is a density matrix.
\end{theorem}
Horodecki et al. \cite{horodecki:separability:1996} showed that this
condition is sufficient for separability in $\C^2\otimes \C^2$ and
$\C^2\otimes \C^3$, but not for other tensor products.  A density
matrix having a positive semidefinite partial transpose is often
referred to as the Peres-Horodecki condition for separability.

In \cite{wu:separable:2006} it was shown that when restricted to zero row sum density matrices, we have a weaker form of the Peres-Horodecki condition that is easier to verify.
\begin{theorem} \label{thm:zerosums}
If a density matrix $A$ with zero row sums is separable, then $A^{PT}$ has zero row sums.
\end{theorem}

\section{normalized Laplacian matrices as density matrices}
For a Laplacian matrix $A$ of a nonempty graph, $\frac{1}{Tr(A)}A$ is symmetric positive semidefinite with trace $1$ and thus can be viewed as a density matrix of a quantum system.  In \cite{braunstein:laplacian_graph:2006} it was shown that a necessary condition for separability of a Laplacian matrix is that the vertex degrees of the graph and its partial transpose are the same for each vertex.  This condition is equivalent to row sums of $A^{PT}$ being $0$. In \cite{wu:separable:2006} it was shown that this condition is also sufficient for separability in $\C^2\otimes \C^q$.  Note that separability of the normalized Laplacian matrix is not invariant under graph isomorphism.  Therefore the vertex numbering is important in determining separability; i.e. we consider labeled graphs.
For labeled graphs of $n$ vertices, there are $2^{\frac{n(n-1)}{2}}$ different Laplacian matrices to consider.  Since the empty graph has trace $0$ and cannot be considered a density matrix, we only need to look at $L(n) = 2^{\frac{n(n-1)}{2}} - 1$ different matrices.

\section{A sufficent condition for separability of normalized Laplacian matrices}

\begin{definition}
A square matrix is {\em line sum symmetric} if the $i$-th column sum is equal to the $i$-th row sum for each $i$.
\end{definition}

\begin{theorem}[\cite{wu:separable:2006}]\label{thm:sufficient}
A normalized Laplacian matrix $A$ is separable in $\C^p\otimes \C^q$ if $A^{i,j}$ in Eq. (\ref{eqn:A}) is line sum symmetric for all $i$,$j$.
\end{theorem}

\begin{figure}[htbp] 
\centerline{
\includegraphics[width=2in]{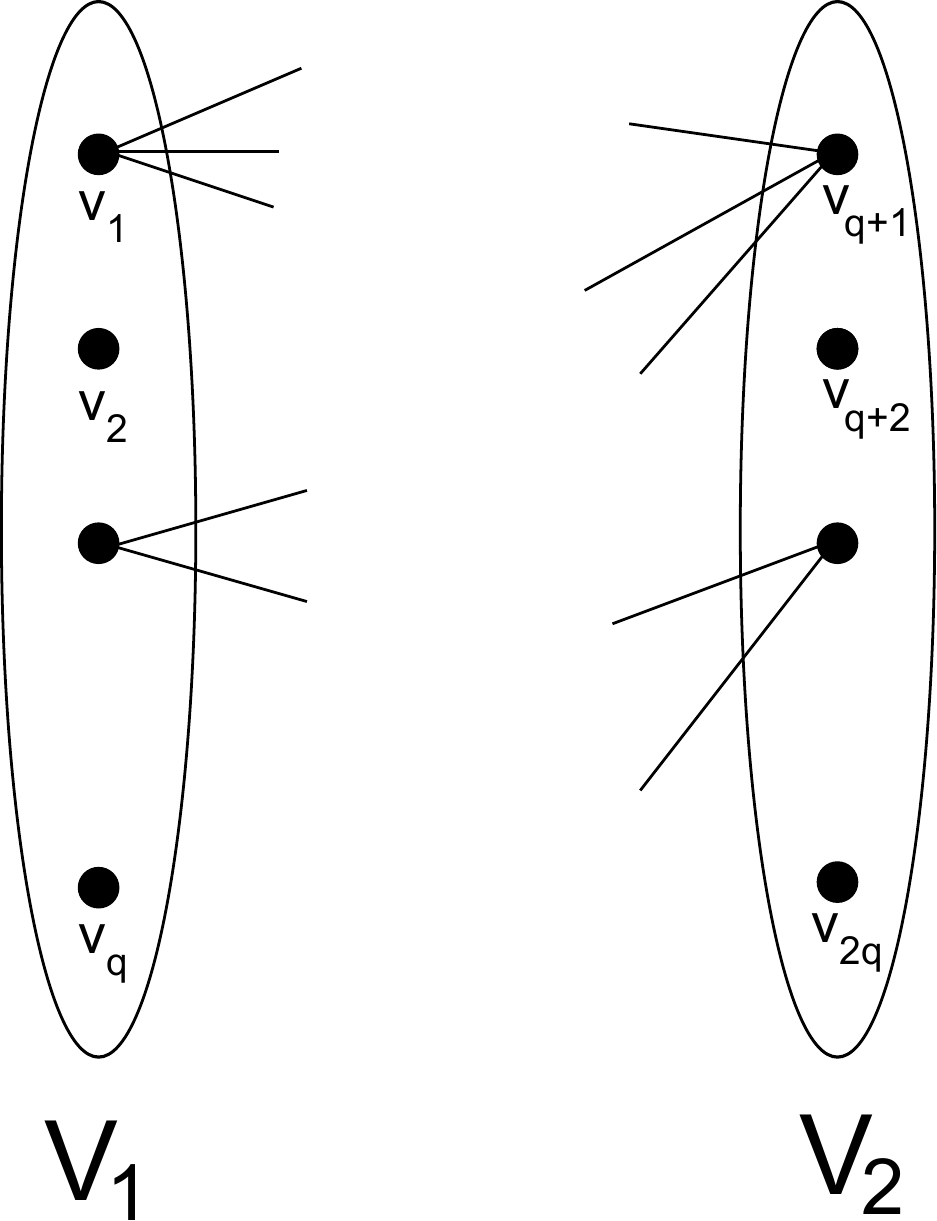}
}
\caption{A sufficient condition for separability for the case $p=2$ is that the number of edges from vertex $v_i$ to $V_2$ is the same as the edges from vertex $v_{q+i}$ to $V_1$ for each $i$.}
\label{fig:sufficient}
\end{figure}

For $V_1$ and $V_2$ disjoint subsets of vertices of a graph, let $e(V_1,V_2)$ denote the number of edges between $V_1$ and $V_2$.
A graphical interpretation of Theorem \ref{thm:sufficient} is that by splitting the $pq$ vertices into $p$ groups $V_i$ of $q$ vertices, where $V_i = \{(i-1)q+1, (i-1)q+2, ..., iq\}$, the normalized Laplacian matrix of a graph ${\cal G}$ is separable in  $\C^p\otimes \C^q$ if for each $j\neq i$ and for each $1\leq m\leq q$, $e(v,V_j) = e(w,V_i)$ where $v$ is the $m$-th vertex in $V_i$ and $w$ is the $m$-th vertex in $V_j$.
This is illustrated in Fig. \ref{fig:sufficient} for the case $p=2$.

\begin{definition} 
Let $L_s(p,q)$ be the number of normalized Laplacian matrices of graphs of $n$ vertices that are separable under $\C^p\otimes \C^q$
where $n = pq$. Let $L_e(p,q)$ be the number of normalized Laplacian matrices of graphs of $n$ vertices that are entangled under $\C^p\otimes \C^q$.  
\end{definition}

It is clear that $L_s(p,q) + L_e(p,q) = L(pq)$.  

\subsection{Upper and lower bounds for $L_s$ and $L_e$}

\begin{definition}
Let ${\cal N}_s(n)$ denote the set of $n$ by $n$ $0$-$1$ matrices that are line sum symmetric.  Let $N_s(n)$ denote the cardinality of the set ${\cal N}_s(n)$.  Let ${\cal N}_e(n)$ denote the set of $n$ by $n$ $0$-$1$ matrices that are not line sum symmetric.  Let $N_e(n)$ denote the cardinality of the set ${\cal N}_e(n)$. 
\end{definition}
Clearly $N_s(n)+N_e(n) = 2^{n^2}$.  The first few values of $N_s(n)$ can be found in \url{https://oeis.org/A229865}.  

We now show how bounds for $L_s$ (and $L_e$) can be derived from $N_s$.
\begin{theorem}
\[ L_s(p,q) \geq 2^\frac{pq(q-1)}{2}N_s(q)^{\frac{p(p-1)}{2}} -1 \]
\end{theorem}
\proof
Let $A$ be the normalized Laplacian matrix of a graph.  Since $A$ is symmetric, in the decomposition in Eq. (\ref{eqn:A}), $A^{i,i}$ is symmetric and $A^{i,j} = (A^{j,i})^T$.  Therefore to apply Theorem \ref{thm:sufficient} we only need to check that $A^{i,j}$ is line sum symmetric for $j> i$.  There are ${\frac{p(p-1)}{2}}$ such submatrices and 
thus $N_s(q)^{\frac{p(p-1)}{2}}$ possible combinations.  The remaining entries in the strictly upper triangular portion of $A$ corresponds to 
$\frac{pq(pq-1)}{2} - \frac{p(p-1)}{2}q^2 = \frac{pq(q-1)}{2}$ elements which corresponds to $2^\frac{pq(q-1)}{2}$ combinations.
This combines to $2^\frac{pq(q-1)}{2}N_s(q)^{\frac{p(p-1)}{2}}$ combinations.  Finally we need to subtract $1$ for the zero matrix corresponding to the empty graph.
\eop

\begin{definition}
Let ${\cal M}_n(i)$ denote the set of symmetric $n$ by $n$ $0-1$ matrices such that
\begin{itemize}
\item There is at least one row with a single $1$.
\item The diagonal entries are $0$
\item There are $2i$ nonzero elements in the matrix.
\end{itemize}
Let $M_n(i)$ denote the cardinality of the set ${\cal M}_n(i)$.
\end{definition}

Some values of $M_n(i)$ are shown in Table \ref{tbl:mni}.  It is clear that ${\cal M}_n(i)$ is the set of adjacency matrices of labeled graphs of $n$ vertices and $i$ edges with at least one vertex of degree $1$ and $M_n(i)$ is the number of such graphs.

\begin{theorem}
$M_n(i) = 0$ if $i > \frac{(n-1)(n-2)}{2} + 1$.  For $j\geq 0$, $n\geq 4+j$,
\[M_n\left(\frac{(n-1)(n-2)}{2} - j + 1\right) = n(n-1)\left(\begin{array}{c}\frac{(n-1)(n-2)}{2}\\j\end{array}\right).\]
For $i\leq 3$, $M_n(i)$ is equal to the number of labeled bipartite graphs with $n$ vertices and $i$ edges.\footnote{See  \url{https://oeis.org/A000217},\url{https://oeis.org/A050534}, \url{https://oeis.org/A053526}.}  In particular, $M_n(1) = \frac{n(n-1)}{2}$, $M_n(2) = \frac{(n+1)n(n-1)(n-2)}{8}$, $M_n(3) = \frac{((n+1)(n+2)+2)n(n-1)(n-2)(n-3)}{48}$. 
\end{theorem}
\proof
Clearly the maximum number of edges in a graph with at least one vertex of degree $1$ is achieved with a single vertex $v$ connected to a vertex $w$ in a clique of $(n-1)$ vertices, and this graph has $\frac{(n-1)(n-2)}{2} + 1$ edges.  If $n\geq 4+j$ and a graph $\cal G$ with $\frac{(n-1)(n-2)}{2} - j + 1$ edges has at least one vertex of degree $1$, then this graph will have a single vertex $v$ connected to a vertex $w$ in a graph ${\cal W}$ consisting of $n-1$ vertices and $\frac{(n-1)(n-2)}{2} - j$ edges, i.e. $\cal W$ is a clique of $n-1$ vertices minus $j$ edges.  Each vertex of $\cal W$ has degree $\geq n-2-j \geq 2$, i.e. $\cal W$ does not include a vertex of degree $1$.  In this case ${\cal G}$ is uniquely defined by the vertices $v$ and $w$ and the graph $\cal W$.  There are $n(n-1)$ such pairs of vertices $(v,w)$ and there are $\left(\begin{array}{c}\frac{(n-1)(n-2)}{2}\\j\end{array}\right)$ labeled graphs with $n-1$ vertices and $\frac{(n-1)(n-2)}{2} - j$ edges.

Let ${\cal G}$ be a graph of $n$ vertices with at least one vertex of degree $1$ and at most $3$ edges.  Since there are no cliques of $3$ vertices, it is a bipartite graph.  Similarly if ${\cal G}$ is a bipartite graph of at most $3$ edges, the absence of cliques of $3$ vertices means that there is a vertex of degree $1$.
\eop

\begin{table}
\begin{center}
\small
\begin{tabular}{c|*{8}r}
$n$ & $M_n(1)$ & $M_n(2)$ & $M_n(3)$ & $M_n(4)$ &$M_n(5)$ & $M_n(6)$ & $M_n(7)$ & $M_n(8)$  \\ \hline\hline
$2$ & $1$ &&&&&&& \\ \hline 
$3$        &$3$ &          $3$ &&&&&& \\\hline
$4$   &        $6$  &        $15$ &         $16$ & $12$ &&&&\\\hline
$5$ &   $10$&          $45$&         $110$&         $195$ &         $210$& $120$ &$20$ &\\\hline
$6$&$15$&$105$&$435$&$1320$&$2841$&$4410$&$4845$&$3360$\\\hline
$7$& $21$      &   $210$  &  $1295$& $5880$ &  $19887$  &  $51954$  &  $106785$ &$171360$\\\hline 
$8$&$28$& $378$& $3220$ &$20265$& $97188$& $369950$ &$1147000$& $2931138$\\\hline
\end{tabular}
\end{center}
\caption{Values of $M_n(i)$.}
\label{tbl:mni}
\end{table}

\begin{theorem}
\[L_e(p,q) \geq \sum_{i=1}^{\frac{(p-1)(p-2)}{2}+1}M_p(i)N_e(q)^iN_s(q)^{\frac{p(p-1)}{2}-i}2^{\frac{pq(q-1)}{2}}\]
\end{theorem}
\proof
Consider the mapping where we replace each submatrix $A^{i,j}$ with $1$ if it is an element of ${\cal N}_e(q)$ and $0$ otherwise.
This results in a $p$ by $p$ $0-1$ matrix $B$.  If $B$ is in ${\cal M}_p(k)$ for some $k$, then 
$A$ has a row of submatrices $A^{i,j}$ that are all line sum symmetric except for one, and this implies that
$A^{PT}$ does not have zero sums and thus $A$ is not separable by Theorem \ref{thm:zerosums}. There are $M_p(k)N_e(q)^kN_s(q)^{\frac{p(p-1)}{2}-k}$ such combinations. As before, there are $\frac{pq(q-1)}{2}$ remaining locations in the strictly upper triangular part of $A$ that is not occupied by $A^{i,j}$ for some $i\neq j$. 
\eop

\begin{corollary}
\[ L_s(p,q) \leq L(pq) - \sum_{i=1}^{\frac{(p-1)(p-2)}{2}+1}M_p(i)N_e(q)^iN_s(q)^{\frac{p(p-1)}{2}-i}2^{\frac{pq(q-1)}{2}} \]
\[ L_e(p,q) \leq L(pq) - 2^\frac{pq(q-1)}{2}N_s(q)^{\frac{p(p-1)}{2}} +1 \]
\end{corollary}

It is easy to show that the upper and lower bound for the case $p=2$ coincide and this corresponds to the fact that the separability (and entanglement) condition is both sufficient and necessary in $\C^2\otimes \C^q$ \cite{wu:separable:2006}.  In particular,

\begin{theorem}
\[ L_s(2,q) = 2^{q(q-1)}N_s(q) - 1\]
\[ L_e(2,q) = 2^{q(q-1)}N_e(q) \]
\end{theorem}
\proof The upper bound for $L_s(2,q)$ is equal to 
\[ L(2q) - N_e(q)2^{q(q-1)} = 2^{q(2q-1)}-1-(2^{q^2}-N_s(q))2^{q(q-1)}  
= N_s(q)2^{q(q-1)}-1 \]
which is also the lower bound for $L_s(2,q)$. \eop

\bibliography{quant,markov,consensus,secure,synch,misc,stability,cml,algebraic_graph,graph_theory,control,optimization,adaptive,top_conjugacy,ckt_theory,math,number_theory,matrices,power,quantum}

\end{document}